\documentclass[english,american,amssymb,amsmath]{revtex4}
\usepackage[T1]{fontenc}
\usepackage[latin9]{inputenc}
\usepackage{amsmath,amsthm}
\usepackage{amssymb}
\usepackage{esint}
\usepackage[all, knot]{}
\makeatletter
\@ifundefined{textcolor}{}
{%
 \definecolor{BLACK}{gray}{0}
 \definecolor{WHITE}{gray}{1}
 \definecolor{RED}{rgb}{1,0,0}
 \definecolor{GREEN}{rgb}{0,1,0}
 \definecolor{BLUE}{rgb}{0,0,1}
 \definecolor{CYAN}{cmyk}{1,0,0,0}
 \definecolor{MAGENTA}{cmyk}{0,1,0,0}
 \definecolor{YELLOW}{cmyk}{0,0,1,0}
 }

\usepackage[all, knot]{xy}

\usepackage{esint}
\usepackage[all, knot]{}
\makeatother

\usepackage{amstext}
\usepackage{amssymb}
\usepackage{babel}

\begin{document}

\title{Higgs potential and confinement in Yang-Mills theory on exotic $\mathbb{R}^4$}

\author{Torsten Asselmeyer-Maluga}

\email{torsten.asselmeyer-maluga@dlr.de}

\affiliation{German Aero Space Center, Rutherfordstr. 2, 12489 Berlin}

\author{Jerzy Kr\'ol}

\email{ iriking@wp.pl}

\affiliation{University of Silesia, Institute of Physics, ul. Uniwesytecka 4,
40-007 Katowice}
\begin{abstract}
We show that pure $SU(2)$ Yang-Mills theory formulated on certain exotic 
$\mathbb{R}^{4}$
from the radial family shows confinement. 
The condensation of magnetic monopoles and the qualitative form of the Higgs 
potential
are derived from the exotic $\mathbb{R}^{4}$, $e$. A relation between
the Higgs potential and inflation is discussed. Then we obtain a formula
for the Higgs mass and discuss a particular smoothness structure so
that the Higgs mass agrees with the experimental value. The singularity
in the effective dual $U(1)$ potential has its cause by the exotic 4-geometry
and agrees with the singularity in the maximal abelian gauge scenario.
We will describe the Yang-Mills theory on $e$ in some limit as the
abelian-projected effective gauge theory on the standard $\mathbb{R}^{4}$.
Similar results can be derived for $SU(3)$ Yang-Mills theory on an
exotic $\mathbb{R}^{4}$ provided dual diagonal effective gauge bosons
propagate in the exotic 4-geometry. 
\end{abstract}


\maketitle

\section{Introduction}

The first indication that Dirac magnetic monopoles might be related
with certain small exotic $\mathbb{R}^{4}$s from the fixed radial
family, appeared in \cite{AsselmeyerKrol2009,AsselmeyerKrol2011}.
The derivation was based on the algebraic agreement between the magnetic
charges of Dirac monopoles and the Godbillon-Vey (GV) classes of certain
foliations generating small smooth exotic $\mathbb{R}^{4}$s grouped
in the radial family. In our recent work \cite{Asselm-Krol-2012g}
we pushed forward the approach and showed that in the radial family
there exists (at least one) exotic $e$ representing the geometry
of the BPS monopoles moduli space, ${\cal M}_{2}^{0}$, where the
Higgs potential vanishes. When Yang-Mills (YM) theory is formulated
on $e$ and when the smooth structure on $e$ is reduced to the standard
one, then Polyakov-'t Hooft BPS monopoles of the 3-dimensional Yang-Mills-Higgs
theory appear as an assignment of the exotic $\mathbb{R}^{4}$, $e$.
This assignment was possible by considering the special foliated topological
limit (FTL) of general relativity (GR) on $e$, so that quasi-modularity
of the metrics became important. However, the geometry of ${\cal M}_{2}^{0}$
is not the exotic geometry of the end of $e$. It is just the 'semi-classical'
approximation where the gravitational Euclidean path integral on $e$
is considered and where ${\cal M}_{2}^{0}$ contributes as gravitational
instanton \cite{Asselm-Krol-2012g}.

In this paper we extend this approach to the general non-zero Higgs
potential emerging from a full exotic 4-geometry on $\mathbb{R}^{4}$.
The precise shape of the Higgs potential is obtained as the Morse
function describing the exotic handle-body structure. In contrast
to our previous work, we will describe the end of the small exotic
$\mathbb{R}^{4}$, $e$, which is an exotic $S^{3}\times_{\theta}\mathbb{R}$.
In this way the BPS condition, hence zero Higgs potential, corresponds
merely to the 4-geometry, i.e. Atiyah-Hitchin gravitational
instanton, whereas the general non-zero Higgs potential corresponds
to the true exotic handle-bodies with non-canceling (smoothly) pairs
of handles. Furthermore, by assuming a connection to inflation and choosing a special exotic $S^{3}\times_{\theta}\mathbb{R}$, we
are able to calculate the Higgs mass which is in a good agreement with
the experimental value.

Thus, given the YM theory on exotic $e$ the Higgs field can be
introduced from this exotic 4-geometry, and the Yang-Mills-Higgs (YMH)
theory with the general Higgs potential can be correctly described. In this way we would have a kind of mechanism for geometric confinement in the $SU(2)$ YM theory
on an exotic $\mathbb{R}^{4}$. Magnetic monopoles and the Higgs field
are now rather ingredients of the exotic 4-geometry than in pure YM on the standard $\mathbb{R}^{4}$. That is why a more direct connection of (twisted) YM $SU(2)$ theory without Higgs field and magnetic monopoles, with exotic 4-geometry should exist.
In this paper we describe this connection in two steps, realizing a geometric confinement in the YM theory on an exotic $\mathbb{R}^{4}$.

1) YM theory is twisted so that it fits with the current and magnetic
monopoles as well Higgs potential from exotic $\mathbb{R}^{4}$. The
twisting is just the abelian (maximal) gauge and is the result of
the asymmetric propagation of gauge fields (gluons). This is the place
where exotic 4-geometry intervenes directly. The propagation relies
on the choice that diagonal dual $U(1)$ fields live on exotic $\mathbb{R}^{4}$
while the electric field is not sensitive to exotic 4-geometry of
the background and propagates in the standard smooth structure of
$\mathbb{R}^{4}$. This asymmetry is the geometric reason for the
abelian projection in YM theory. There exist various twisting versions
of YM theory and we discuss here two of them: The maximal abelian
gauge and abelian projected effective gauge theory (APEGT) \cite{Kondo-1997,Kondo-2011}
theories. Both theories show confinement.

2) However, the exotic 4-geometry of $e$ carries also the mechanism
for breaking the dual magnetic $U(1)$ symmetry (and the Bianchi identity).
This is the singularity in the magnetic $U(1)$ field due to the change
of the smoothness from exotic to the standard one. The fixed radial
family introduces the magnetic current into the YM theory which is
nothing but the nontrivial Godbillon-Vey class of the associated codimension-1
foliation. The non-zero magnetic current, in turn, via the condensation
of magnetic monopoles, generates the confinement in the YM theory.

This 2-step process indicates the very fundamental role played by
exotic $\mathbb{R}^{4}$s, in particular the non-trivial 4-geometry
which becomes an important ingredient to explain the confining/deconfining
change of phases in YM theory.

We close the paper with the short discussion of the proposed geometric
scenario for the confinement in $SU(2)$ theory on $e$. Still, a
more thorough understanding of the physical meaning of the geometric
asymmetry in the gluon propagation is needed. This problem, along
with the analysis of the physical case of $SU(3)$ QCD on exotic $\mathbb{R}^{4}$,
will be presented separately.

\section{3-d BPS magnetic monopoles and Higgs field in YMH theory from exotic
$\mathbb{R}^{4}$}

\label{2} In this section we recapitulate some of the results obtained
in \cite{Asselm-Krol-2012g}. The 4-d $SU(2)$ YMH theory on $\mathbb{R}^{4}$
with the Minkowski metric, is given by the density of the Lagrangian
\cite{MM-1998-PHD}: 
\begin{equation}
{\cal L}=-\frac{1}{2}{\rm Tr}(F_{\mu\nu}F^{\mu\nu})+{\rm Tr}(D^{\mu}\Phi D_{\mu}\Phi)-U(\Phi)=-\frac{1}{4}F_{\mu\nu}^{a}F^{a\mu\nu}+\frac{1}{2}D^{\mu}\Phi^{a}D_{\mu}\Phi^{a}-U(\Phi^{a})\label{20}
\end{equation}
where $U(\Phi)=-\frac{\lambda}{2}({\rm Tr}(\Phi^{2})-v^{2})^{2}=\frac{\lambda}{4}(\Phi^{a}\Phi^{a}-v^{2})^{2}$
is the gauge-invariant potential and $a$ is the internal index in
the $su(2)$ algebra. One sees that the spectrum of this theory contains
three massless vector bosons and a massive scalar Higgs field. Defining
the fields $E^{ai}=F^{ai0}$ and $B^{ai}=-\frac{1}{2}\epsilon^{ijk}F_{ajk}$
in the usual way , the energy of the theory reads: 
\begin{equation}
{\cal E}=\frac{1}{2}\{E_{i}^{a}E_{i}^{a}+B_{i}^{a}B_{i}^{a}+D_{0}\Phi^{a}D_{0}\Phi^{a}+D_{i}\Phi D_{i}\Phi\}+U(\Phi)\,,\label{En}
\end{equation}
so that the vacuum configuration is given by $A_{\mu}=0$ and $|\Phi|=v={\rm const}$
along the gauge transformations of this theory. Given the vacuum as
$A_{\mu}^{v}=0$, $\Phi^{v}=(0,0,v)$ and choosing the gauge so that
$\phi=(0,0,\phi_{3})$ and $\Phi=\Phi^{v}+\phi$, the Lagrangian of
the YMH theory in these new fields, reads: 
\begin{equation}
{\cal L}=-\frac{1}{4}F_{\mu\nu}^{a}F^{a\mu\nu}+\frac{(ev)^{2}}{2}A^{1\mu}A_{\mu}^{1}+\frac{(ev)^{2}}{2}A^{2\mu}A_{\mu}^{2}+\frac{1}{2}\partial^{\mu}\phi_{3}\partial_{\mu}\phi_{3}-\frac{1}{2}\lambda v^{2}(\phi_{3})^{2}\,.
\end{equation}
The spectrum now contains two massive $A_{\mu}^{1},A_{\mu}^{2}$ and
one massless $A_{\mu}^{3}$ vector bosons and the massive $\phi_{3}$
Higgs field with the classical mass $v\sqrt{\lambda}$. In this reformulated
YMH theory one embeds the electromagnetism such that magnetic monopoles
appear. This is performed via the gauge-invariant electromagnetic
tensor found by 't Hooft: 
\[
f_{\mu\nu}={\rm Tr}(\hat{\Phi}F_{\mu\nu})-\frac{1}{e}{\rm Tr}(\hat{\Phi}[D_{\mu}\hat{\Phi},D_{\nu}\hat{\Phi}])=\partial_{\mu}(A_{\nu}^{a}\hat{\Phi}^{a})-\partial_{\nu}(A_{\mu}^{a}\hat{\Phi}^{a})-\frac{1}{e}\epsilon^{abc}\hat{\Phi}^{a}\partial_{\mu}\hat{\Phi}^{b}\partial_{\nu}\hat{\Phi}^{c}
\]
where $\hat{\Phi}=\frac{\Phi}{{\rm Tr}(\Phi^{2})}$. New electric
and magnetic fields as well as electric, $j_{\mu}^{e}$, and magnetic,
$j_{\mu}$, currents, thus read: 
\begin{equation}
\begin{array}{c}
E^{i}=f^{i0},B^{i}=-\frac{1}{2}\epsilon^{ijk}f_{jk}\\[5pt]
j_{\mu}^{e}=-\partial_{\nu}f_{\nu\mu},\, j_{\mu}=\partial_{\nu}^{\star}f_{\nu\mu}=-\frac{1}{2}\epsilon_{\mu\nu\rho\sigma}\partial^{\nu}f^{\rho\sigma}=-\frac{1}{2e}\epsilon_{\mu\nu\rho\sigma}\epsilon^{abc}\partial_{\nu}\hat{\Phi}^{a}\partial_{\rho}\hat{\Phi}^{b}\partial_{\sigma}\hat{\Phi}^{c}
\end{array}\label{YMH-1}
\end{equation}
where it is seen that $\partial^{\mu}j_{\mu}=0$. This electromagnetic
theory exhibits electric-magnetic duality: 
\[
\begin{array}{c}
j_{0}=\nabla\cdot B,\;\;\; j_{i}=-(\nabla\times E)_{i}-\partial_{t}B_{i}\\[5pt]
j_{0}^{e}=-\nabla\cdot E,\;\;\; j_{i}^{e}=(\nabla\times B)_{i}-\partial_{t}E_{i}\,.
\end{array}
\]
The equations of motion derived from the Lagrangian of the YMH theory
(YMH eqs.), read: 
\[
D_{\mu}F^{\mu\nu}=e[\Phi,D^{\nu}\Phi],\;\;\; D^{\mu}D_{\mu}\Phi=-\lambda\Phi{\rm Tr}(\Phi^{2}-v^{2})\,.
\]
In order to have finite energy configurations one has to have $U(\Phi)\underset{x\to\infty}{\rightarrow}0$
from (\ref{En}) so that $\Phi^{a}\Phi^{a}\underset{x\to\infty}{\rightarrow}v^{2}$.
Thus, the topological charge, the winding number $N$ of the 2-sphere
by the 2-sphere as element of the homotopy group $\pi_{2}(S^{2})=\mathbb{Z}$,
is related to the total magnetic charge $g$ of the field configuration:
\[
g=\frac{4\pi N}{e}\,.
\]
However, to solve the YMH equations in full generality is the formidable
task, and because of that one considers the special limit, namely
the BPS limit, where $U(\Phi)$ is zero in the YMH Lagrangian. Moreover,
the BPS and static configurations are considered. In this way, the
YMH theory reduces to 3-dimensions and the space of solutions is tractable
and can be explicitly presented. This exactly solvable, though reduced
to 3-d, case allows one to relate the geometry of the moduli space
${\cal M}_{2}^{0}$ of the $k=2$ magnetic monopoles, with some exotic
4-geometry on $\mathbb{R}^{4}$. All small exotic $\mathbb{R}^{4}$s
are considered as being grouped in the fixed radial family of such
structures which happens to be crucial in derivation of the relation
\cite{Asselm-Krol-2012g}.

The above mentioned 3-d reduction of the 4-d YMH theory on Minkowski
4-space, can be also seen as the reduction of the 4-d pure YM theory
on the Euclidean $\mathbb{R}^{4}$. The Higgs field appears in 3-d
YMH theory from this 4-d YM theory though with the vanishing Higgs
potential. Moreover, the self-duality of 4-d Euclidean pure YM theory
enforces the theory to fulfill the BPS condition after reduction to
3-d. We would like to find a similar appearance of the Higgs field
and magnetic monopoles also for a general Higgs potential. This potential
is attempted to be obtained from an exotic $\mathbb{R}^{4}$, which the YM theory
is formulated on.

First, let us recall how the 4-d to 3-d reduction looks like for the
YM theory on Euclidean $\mathbb{R}^{4}$. The pure Euclidean $SU(2)$
YM theory on $\mathbb{R}^{4}$ is given by the action 
\begin{equation}
S=-\frac{1}{2}\int d^{4}x\,{\rm Tr}(F_{\mu\nu}^{a}F^{a\mu\nu})\label{YM-4}
\end{equation}
where as usual $F_{\mu\nu}^{a}=\partial_{\mu}A_{\nu}^{a}-\partial_{\nu}A_{\mu}^{a}+f_{\beta\gamma}^{a}A_{\mu}^{\beta}A_{\nu}^{\gamma}$
with structure constants $f_{\beta\gamma}^{a}$ of $SU(2)$ and the
vector potential $A_{\mu}^{a},\mu=1,2,3,4$ takes values in the $su(2)$
algebra. The dual tensor $^{\star}F$ to $F$ is defined by $^{\star}F_{\mu\nu}^{a}=\epsilon^{\mu\nu\rho\sigma}F_{\rho\sigma}^{a}$.
The equations of motion (EOM) derived from (\ref{YM-4}) in the component-free
notation, read: 
\[
D_{A}\, F=0=D_{A}{^{\star}F}
\]
where $D_{A}=d+A$ is the covariant exterior derivative (depending
on the connection $A$). The (anti-)self-duality condition now reads
$F_{\mu\nu}^{a}=\mp^{\star}F_{\mu\nu}^{a}$ and the connection $A$
realizing this equation, solves automatically the above YM EOM since
they reduce to the Bianchi equations.

Suppose now that $A_{4}=\phi$ such that $A=A_{1}dx_{1}+A_{2}dx_{2}+A_{3}dx_{3}+\phi dx_{4}$
where $A_{i},i=1,2,3$ and $\phi$ are $su(2)$-algebra valued functions
on $\mathbb{R}^{3}$. Also, the YM EOM (\ref{YM-4}) are invariant
with respect to the translations in the $x_{4}$-directions, meaning
that the transformed and the original configurations are equivalent
up to a gauge. The Euclidean YM Lagrangian for this reduced YM theory
reads: 
\begin{equation}
{\cal L}=-\frac{1}{2}{\rm Tr}(F_{\mu\nu}F^{\mu\nu})+{\rm Tr}(D^{\mu}\phi D_{\mu}\phi)=-\frac{1}{4}F_{\mu\nu}^{a}F^{a\mu\nu}+\frac{1}{2}D^{\mu}\phi^{a}D_{\mu}\phi^{a}\,.
\end{equation}
where $F$ and $D_{A}$ are now defined with respect to the connection
$A_{1}dx_{1}+A_{2}dx_{2}+A_{3}dx_{3}$ with the translational invariance
in $x_{4}$ of the expressions. In this way we have a 3-d YMH theory
with $\lambda=0$, so that the Higgs potential vanishes. The Bogomolny
equations in this setting give: 
\[
B_{i}=D_{i}\phi=D_{i}A_{4}=E_{i}
\]
which means that the connection is self-dual. Thus, finally we have
a 3-d YMH theory with vanishing Higgs potential fulfilling the Bogomolny
equations, i.e. the static BPS monopoles configurations are described
here, and this theory is the dimensional reduction of the pure self-dual
Euclidean 4-d YM theory (without any Higgs field) \cite{AH-1988,MM-1998-PHD}.
The quantum dynamics of these BPS monopoles, hence 4-d case, is partially
restored (in the low energy limit and for slow motions) by considering
the geodesic approximation on the space of moduli of the static BPS
monopoles \cite{Stuart-1994}. On the other hand, the geometry of
the space of the moduli ${\cal M}_{2}^{0}$ of the $k=2$ magnetic
monopoles is related with the exotic 4-geometry on some fake 4-space
from the radial family \cite{Asselm-Krol-2012g}. It was performed
by considering the special \emph{foliated topological limit} (FTL)
of general relativity on exotic $\mathbb{R}^{4}$. In this limit quasi-modular,
i.e. depending on the Eisenstein 2-nd series $E_{2}$, expressions
become dominant and so, the geometry of ${\cal M}_{2}^{0}$ does.
However, it is still an approximation to the true exotic geometry
of the end of exotic $\mathbb{R}^{4}$, $e$. It is conjectured that
precisely ${\cal M}_{2}^{0}$, being the gravitational instanton,
gives the dominant contribution to the Euclidean path integral on
the exotic end of $e$ \cite{Asselm-Krol-2012g}.

Thus the assignment of $k=2$ magnetic monopoles to exotic $\mathbb{R}^{4}$
is rather effective and approximative. On the one hand this is due
to the absence of Higgs potential in the BPS limit. On the other hand
the reason is the argumentation based on semi-classical path integral.
To approach confinement in such setting we would like to have rather
the general Higgs potential and the reduction to the dual abelian
Higgs theory. In the next section we are going to show that the shape
of general Higgs potential follows from the exotic handle-body topology
of $e$.

\section{General Higgs potential from open exotic 4-geometry}

\label{3}

Let us consider the end of a small exotic $\mathbb{R}^{4}$, i.e.
a sequence of compact (codim-0) subspaces $K_{1}\subset K_{2}\subset K_{3}\subset\ldots$
leading to a sequence $U_{1}\subset U_{2}\subset U_{3}\subset\ldots$
of complements $U_{i}=\mathbb{R}^{4}\setminus K_{i}$. The end of
every topological $\mathbb{R}^{4}$ is topologically $S^{3}\times\mathbb{R}$.
Furthermore, the end of an exotic%
\footnote{We exclude the case of an exotic $\mathbb{R}^{4}$ which produces
an exotic $S^{4}$ by compactification. No example of an exotic $S^{4}$
is known.%
} $\mathbb{R}^{4}$ is an exotic $S^{3}\times\mathbb{R}$. In the following
we will discuss the appearance of the Higgs potential from the topology
of the exotic $S^{3}\times_{\theta}\mathbb{R}$ as end of a small
exotic $\mathbb{R}^{4}$. In contrast to our previous work where the
exotic $S^{3}\times_{\theta}\mathbb{R}$ was simply constructed from
the homology 3-sphere $\Sigma$ appearing as the smooth cross-section
\cite{Fre:79,AsselmKrol-2013b} we need now the full power of Casson
handles. But surprisingly the result is very similar. At first we
state a known result \cite{GomSti:1999,Ganz2000} that $n$th untwisted
Whitehead double of the pretzel knot (3,3,-3) (or the knot $9_{46}$
in Rolfson notation) separates the end ($S^{3}\times_{\theta}\mathbb{R}$)
from the small exotic $\mathbb{R}^{4}$. But then we have to consider
only the Casson handle $CH$ in $S^{3}\times_{\theta}\mathbb{R}$
coming from a non-canceling pair of 1-/2-handles (see below for an
explanation). By definition, this $CH$ is exotic, i.e. the attaching
circle is not smoothly slice. Bizaca (see for instance Corollary 0.3
in \cite{BizGom:96}) constructed a family of Casson handles fulfilling
this property. The attaching circle is bounded by a knot. The idea
of the proof is the construction of a knot so that any (finite) n-fold
untwisted Whitehead double of this knot does not bound a smooth disk.
The result is the right-handed (or positive) trefoil knot. Following
\cite{Ganz2000}, we constructed a smooth section in $S^{3}\times_{\theta}\mathbb{R}$
by Dehn surgery in dependence on the framing of the attaching circle.
For the framing $+1$ we obtain the Poincare homology 3-sphere and
a sequence of homology 3-spheres as Whitehead doubles of the trefoil
knot. In particular, every other knot smoothly concordant to the trefoil
knot (like the knot $8_{10}$ see below) can be also used. But one
thing remains: we obtain a (homology) cobordism (as a 4-manifold)
in the exotic $S^{3}\times_{\theta}\mathbb{R}$ between $S^{3}$ and
a homology 3-sphere $\Sigma$ (not bounding a contractable, smooth
4-manifold).

The Morse theory associated with the manifold $\Sigma$ is determined
by considering a scalar function $f:\,\Sigma\to[0,1]$ over $\Sigma$.
The number of critical points of the Morse-function $f$ are related
to the dimension of the homology-groups of $\Sigma$. Let us consider
two Morse-functions $\psi:\, S^{3}\to\mathbb{R}$ and $\Psi:\,\Sigma\to\mathbb{R}$.
In the context of cosmological models one can consider them as corresponding
to two cosmic states $S^{3}$ and $\Sigma$ in the exotic $S^{3}\times_{\theta}\mathbb{R}$
model where $\Sigma$ is the homology 3-sphere constructed from the
knot $8_{10}$ (see \cite{AsselmKrol-2012i} for an explanation of
this choice). The spacetime and in particular its smoothness structure
is represented by the homology cobordism $\phi:\,\Sigma\to S^{3}$
(a cobordism preserving the homology groups) between these states.
This mapping $\phi$ factorizes the Morse-function $\Psi$ on $\Sigma$
in two maps $\Psi=\psi\circ\phi$, i.e. the diagram 
\begin{eqnarray}
\Sigma & \stackrel{\Psi}{\longrightarrow} & \mathbb{R}\nonumber \\
\phi\downarrow & \circlearrowright & \updownarrow id\label{eq:commuting-diagram}\\
S^{3} & \stackrel{\psi}{\longrightarrow} & \mathbb{R}\nonumber 
\end{eqnarray}
commutes. The new map $\psi$ is a representation of $\Psi$ on $S^{3}$.
On the 3-sphere there exist two critical points: one maximum and one
minimum. That means that the cobordism $M$ possesses the homology
groups $H_{0}(M)=H_{3}(M)=\mathbb{Z}$ of a homology 3-sphere. The
evaluation of the exact sequence of the pair $(M,\,\partial M)$ shows
that $H_{0}(M)$ is generated by one of the boundary components, e.g.
by $H_{0}(S^{3})$ and $H_{3}(M)$ by the other one, i.e. $H_{3}(\Sigma)$.
The transition $y=\phi(x)$ represented by $M$ maps the Morse function
$\psi(y)=||y||^{2}$ on $S^{3}$ to the Morse function $\Psi(x)=||\phi(x)||^{2}$.

A field-theoretic discussion of the Morse-theoretic result starts
with the following simple idea: The 3-sphere is isomorphic to $SU(2)$
and so one can define a formal group-operation on $S^{3}$. This makes
the map $\phi$ to a $SU(2)$-valued scalar-field over $\Sigma$.
Witten has derived the Morse-theory from the field-theoretic properties
\cite{Wit:82a}. His ansatz gives a field-theoretic construction of
the dynamics which gives a ``tunneling path'' between two critical
points. In our case, the field of the Witten-construction is the $SU(2)$-scalar
$\phi$ over the homology 3-sphere $\Sigma$. Using his $\sigma$-model
we get a field theory with the Lagrange density
\[
\mathcal{L}_{\xi}=D_{\mu}\phi\cdot D^{\mu}\phi+\phi\cdot\phi\,,
\]
with the covariant derivation $D_{\mu}$ and a representation dependent
product $\cdot$ of the group $SU(2)$. Adding the Einstein-Hilbert-action
we obtain the well known model of chaotic inflation
\begin{equation}
\mathcal{L}=\frac{1}{\kappa}R+D_{\mu}\phi\cdot D^{\mu}\phi+\frac{\rho_{G}}{2}\phi\cdot\phi\:,\label{eq:LagrangeInflation}
\end{equation}
so that the topological transition $S^{3}\to\Sigma$ of the cosmological
model can be interpreted as the inflation of the cosmos \cite{AsselmKrol-2013b}
The factor $\rho_{G}$ can be interpreted as the curvature of the
homology 3-sphere $\Sigma$ represented as energy density. The Lagrangian
(\ref{eq:LagrangeInflation}) is only valid, if the cobordism is a
smooth 4-manifold leading to a smooth transition $S^{3}\to\Sigma$.
But in the construction of the exotic $S^{3}\times_{\theta}\mathbb{R}$,
the end of an exotic $\mathbb{R}^{4}$, one needs a non-smooth cobordism
together with a Casson handle reflecting the exoticness of $S^{3}\times_{\theta}\mathbb{R}$.
Casson handles are also responsible for the exotic smoothness on Euclidean
$\mathbb{R}^{4}$ which is our main concern in this paper. Namely,
consider the construction of a homology cobordism, like $\phi:\,\Sigma\to S^{3}$,
where Casson handles have to appear. The problem is the introduction
of the 2-/3-handle pair. The existence of the exotic smoothness structure
forbids the smooth cancellation of the handle pairs so that the infinite
layer structure of the Casson handle appears. The generic case is
given by the following model. Glue a 1-/2-handle pair (dual to a 2-/3-handle
pair) to a 0-handle. The 0-handle is modeled by a minimum in Morse
theory. But the 1-/2-handle pair generates a maximum and a minimum
which can be canceled.

Exotic smoothness forbids this cancellation and we obtain an extra
pair of maximum/minimum. In the notation above by using the commutative
diagram (\ref{eq:commuting-diagram}), we can model a Morse function
with two minima and one maxima by the Morse function
\begin{equation}
\psi(y)=||y||^{4}-||y||^{2}\label{eq:morse-Higgs-potential}
\end{equation}
on $\Sigma$. We describe the transition by a $SU(2)$-valued scalar
field $\phi:\Sigma\to SU(2)$ and obtain
\begin{equation}
\mathcal{L}=\frac{1}{\kappa}R+D_{\mu}\phi\cdot D^{\mu}\phi+\frac{\rho_{G}}{2}(\phi^{4}-\phi^{2})\:,\label{eq:Lagrangian-Higgs-potential}
\end{equation}
as Lagrangian. By using Cerf theory \cite{Cerf1970}, the choice of
the function (\ref{eq:morse-Higgs-potential}) is generic which we
will explain now. The Morse function of the cobordism can be interpreted
as a one-parameter family of Morse functions. Then, there three types
of functions: the usual Morse function (as sum/difference of quadratic
terms), the birth-death point as function $x^{3}$ plus a Morse function
and the dovetail function $x^{4}$ plus a Morse function. Because
of the 2/3- handle pair (codimension-2 case), we have to choose the
dovetail (having a codimension-2 critical point). The unfolding (the
one-parameter family) is $x^{4}\pm tx^{2}$ (up to a linear term and
a Morse function). Therefore the function (\ref{eq:morse-Higgs-potential})
is generic, i.e. every smooth deformation of the spaces in the diagram
(\ref{eq:commuting-diagram}) do not change the form of this function.
This unfolding of the function (\ref{eq:morse-Higgs-potential}) has
also a direct physical background: the deformation parameter can be
interpreted as the mass of the Higgs field
\[
\mathcal{L}=\frac{1}{\kappa}R+D_{\mu}\phi\cdot D^{\mu}\phi+\frac{\rho_{G}}{2}(\phi^{4}-\phi^{2})+\frac{M_{H}}{2}\phi^{2}\:.
\]
But then by a careful study of the transition $S^{3}\to\Sigma$, we
should be able to predict the mass of the Higgs field. In \cite{AsselmKrol-2013b},
we studied this process and discussed its relation to inflation. In
particular we obtained the scaling (or expansion) induced by this
process to be
\begin{equation}
a=a_{0}\cdot\exp\left(\frac{3\cdot vol(\Sigma)}{2\cdot CS(\Sigma)}\right)=L_{P}\cdot\exp\left(\frac{3\cdot vol(\Sigma)}{2\cdot CS(\Sigma)}\right)\label{eq:scaling-inflation}
\end{equation}
where we assume a 3-sphere of radius $L_{P}$ (Planck length) at the
starting point, $vol(\Sigma)$, $CS(\Sigma)$ are the volume and Chern-Simons
functional, respectively, which are topological invariants of the
hyperbolic homology 3-sphere $\Sigma$. Therefore, we have to look
for an expression of the mass which depends on the length scale. Lets
start with the Planck mass
\[
m_{P}=\sqrt{\frac{hc}{G}}
\]
and made the following manipulations
\[
m_{P}=\sqrt[6]{\frac{h^{3}c^{3}}{G^{3}}}=\sqrt[3]{\frac{h^{2}}{G}\cdot\sqrt{\frac{c^{3}}{hG}}}=\sqrt[3]{\frac{h^{2}}{L_{P}\cdot G}}
\]
to obtain a relation between length ($L_{P}$) and mass ($m_{P}$)
(without using the Compton wave length). This relation can be generalized
to 
\[
M(L)=\sqrt[3]{\frac{h^{2}}{L\cdot G}}
\]
a length dependent mass scale. Together with (\ref{eq:scaling-inflation})
we obtain the formula
\[
M=\sqrt{\frac{hc}{G}}\cdot\exp\left(-\frac{vol(\Sigma)}{2\cdot CS(\Sigma)}\right)
\]
Therefore, for a $+8$ Dehn-surgery with a special cusp (generating
a geodesic of minimal length $0.3531$), we obtain a homology 3-sphere
$\tilde{\Sigma}(8_{10})$ with 
\begin{eqnarray*}
vol(\tilde{\Sigma}(8_{10})) & = & 5,902827...\\
CS(\tilde{\Sigma}(8_{10})) & = & 0.07546...
\end{eqnarray*}
and calculate the mass to
\[
M_{H}\approx126\, GeV
\]
This result is not totally surprising, because there is an infinity
of homology 3-spheres and it least one value should fit. The inflation
process with theses values (putted into formula (\ref{eq:scaling-inflation}))
leads to around 117 of e-folds much larger than the minimal required
value of 60 e-folds. Therefore if our assumption about a relation
between inflation and the Higgs boson is correct, then one has to
found experimental hints of an inflation with 117 e-folds of expansion. 

This model is quite general, applicable also to the case of small
exotic smooth $\mathbb{R}^{4}$ where Casson handles can not be smoothly
reduced to the 2-handles. Turning to the $su(2)$ Lie algebra valued
field $\phi=\phi^{a}\cdot T^{a}$ and assuming the adjoint representation
for such $\phi$ and taking $\phi^{a}\phi^{a}=v^{2}\neq0$, one recovers,
up to the additive and multiplicative constants, the appearance of
the general shape of the Higgs potential as in (\ref{20}). That is
why the Higgs potential and field can be modeled by a Morse function
on Casson handles of exotic $\mathbb{R}^{4}$, given a $SU(2)$ YM
theory on it. This is the smooth topological indication on magnetic
monopoles emerging from exotic 4-geometry on $\mathbb{R}^{4}$ which
complements, and extends over non-BPS case, the static approach via
Atiyah-Hitchin moduli space. At this stage, to avoid complications
from the curvature of exotic $\mathbb{R}^{4}$, one considers the
YM theory as on the flat $\mathbb{R}^{4}$, whereas the impact of
exoticness is being the appearance of the Higgs potential from the
Morse function. However, there exists yet another aspect of exotic
smoothness which complements these just discussed and which appears
as important for color confinement in $SU(2)$ YM theories when on
exotic $\mathbb{R}^{4}$. This is the smooth differentiability of
some merely continuous functions on the standard $\mathbb{R}^{4}$.
This is the topic of the next section.

\section{The mechanism for geometric confinement and the dual Meissner effect}

Now we want to show that confinement in YM theories can be generated
by the non-trivial geometry of the background where the theory is
formulated. Here, our concern is the pure $SU(2)$ YM on some exotic
$\mathbb{R}^{4}$, $e$. Note that the topology of $e$ is the same
as $\mathbb{R}^{4}$ but its smoothness structure differs. Even though
exotic $\mathbb{R}^{4}$s are non-flat Riemannian 4-manifolds, the
direct action of their curvature is neglected, whereas various results
derived from $e$ become important. In the following we will make
use of: i) Magnetic monopoles appearing from exotic 4-geometry in
some limit of general relativity (Sec. \ref{2}); ii) the scalar function,
the Higgs potential, originated from the smooth topology of $e$ which is
the Morse function (Sec. \ref{3}); iii) the appearance of the dominant
abelian phase in pure $SU(2)$ YM on the standard $\mathbb{R}^{4}$,
i.e. the abelian projected effective gauge theory (APEGT) \cite{Kondo-1997,Kondo-2011},
which agrees with the action of the Higgs from (ii) and the condensation
of magnetic monopoles from (i) above.

Let us describe briefly the mechanism of confinement of electrically
charged particles ('quarks') via the condensation of monopoles as in
the dual Meissner effect. We begin with the 4-d relativistic generalization
of the Landau-Ginzburg (GL) Lagrangian 
\begin{equation}
S=-\int d^{4}x\left(\frac{1}{4}F_{\mu\nu}F^{\mu\nu}-|(\partial_{\mu}+ieA_{\mu})\phi|^{2}+\frac{\lambda}{4}(\phi\phi^{\star}-v^{2})^{2}\right)\,.\label{LG-1}
\end{equation}
This is an abelian theory with $U(1)$ gauge field and the complex
Higgs field $\phi$. In the dual abelian Higgs theory $\phi$ represents
a magnetic charge condensate and $F_{\mu\nu}$ is replaced by the
dual $\overline{F}_{\mu\nu}$ where the covariant derivative reads:
$D_{\mu}=\partial_{\mu}+igB_{\mu}$ and $g$ is the magnetic charge.
$\overline{F}_{\mu\nu}=\partial_{\mu}B_{\nu}-\partial_{\nu}B_{\nu}+\overline{G}_{\mu\nu}$
where $\overline{G}_{\mu\nu}$ is the usual string term allowing a
non-zero electric current $\partial_{\mu}G^{\mu\nu}=j^{\nu}$ and
electric charges $e$. The Lagrangian is rewritten as: 
\begin{equation}
S[B,\phi,\phi^{\star}]=-\int d^{4}x\left(\frac{1}{4}\overline{F}_{\mu\nu}\overline{F}^{\mu\nu}-\frac{1}{2}(D_{\mu}\phi)(D_{\mu}\phi)^{\star}+\frac{\lambda}{4}(\phi\phi^{\star}-v^{2})^{2}\right)\,.\label{LG-2}
\end{equation}
This action is invariant with respect to the following abelian gauge
transformations: 
\[
B_{\mu}\to B_{\mu}+\partial\beta\,,\;\;\phi\to e^{-ig\beta}\phi\,.
\]
Rewriting $\phi$ as: $\phi(x)=S(x)e^{ig\psi(x)}$, where $S(x)$
and $\psi(x)$ are real, the action reads: 
\begin{equation}
S[B,\psi,S]=\int d^{4}x\left(-\frac{1}{2}\overline{F}^{2}+\frac{g^{2}S^{2}}{2}(B+\partial\psi)^{2}+\frac{1}{2}(\partial S)^{2}-\frac{1}{4}\lambda(S^{2}-v^{2})^{2}\right)\,.\label{LG-3}
\end{equation}
The ground state is now achieved at $S=v$, so that the field $S$
describes a 'Higgs' particle with mass $m_{H}=2v\sqrt{\lambda}$ and
the gauge boson acquires mass $m_{B}=gv$.

Noticing the gauge symmetries of the action (\ref{LG-3}) which read:
\[
B\to B+\partial\beta\,,\;\; S\to S\,,\;\;\psi\to\psi-\beta\,,
\]
and rewriting $S$ as $S=v+s$ the equation of motion for $\tilde{B}_{\mu}=B_{\mu}+\partial\psi$
at the lowest order in the interactions with $s$, becomes: 
\[
\partial^{\mu}\overline{F}_{\mu\nu}+g^{2}v^{2}\tilde{B}_{\nu}=0\,.
\]
This is equivalent to 
\[
(\square+m_{B}^{2})\tilde{B}_{\mu}=0\,,\;\;\partial^{\mu}\tilde{B}_{\mu}=0\,.
\]
In this way it is explicitly seen that the $\tilde{B}$ field (photon)
acquires mass $m_{B}$, hence the field has a finite penetration depth
into the dual superconducting medium which is defined as the inverse
of the photon mass $\Lambda=\frac{1}{m_{B}}$. Another characteristic
is the correlation length which is $L=\frac{1}{m_{H}}$. Magnetic
monopoles can condense and the electric field is expounded from the
superconductor and in case of a 2-nd kind (dual) superconductor ($\Lambda>L$),
the electric flux is quantized and forced to form (dual) Abrikosov
lines, with the potential proportional to the distance between electric
charges. This is more-or-less the idea of the confinement of electric
charges via the magnetic monopoles condensation and the dual Meissner
effect in the abelian YMH theory, through the action of the Higgs
scalar. There are Wilson loop observables detecting the confining
phase, i.e. one shows that the area law, $W(C)\sim e^{-{\rm Area}(C)}$,
is fulfilled for these operators $W(C)=\left<Pe^{i\oint_{C}dx^{\mu}B_{\mu}}\right>$.
The potential, proportional to the separation of electric charges,
follows. The contour $C$ can be chosen as rectangular with 2 parallel
edges one positively oriented with time (charged particle) and one
in the reverse direction (antiparticle).

However, QCD, or our pure $SU(2)$ YM case on Euclidean $\mathbb{R}^{4}$
given by the Lagrangian (\ref{YM-4}), does not contain any scalar
Higgs field and is not any abelian theory. Hence, the emergence of
the Higgs field and its potential from exotic 4-geometry is a way
of introducing them into the YM theory. The pure YM is also deformed
to agree now with the effects of an exotic $\mathbb{R}^{4}$ and Higgs
from it. There are alternative mechanisms for the confinement in non-abelian
theories without any scalar Higgs field included directly into the
YM theory, like maximal abelian gauge (MAG) or abelian gauge projection.
One shows that the choice of this abelian projection indeed gives
rise to the effect of the confinement, but the procedure seemed to
be gauge dependent. As shown by Kondo and his collaborators in a series
of papers, the gauge independent mechanism for confinement in $SU(2)$
and $SU(3)$ YM theories emerges naturally when turning to the effective
theory - APEGT. Also, as shown in Ref. \cite{DiGiacomo-2010} the
MAG method can be formulated in a gauge independent way by making
use of the non-abelian Bianchi identity. Thus, in the case of YM theory
on an exotic $\mathbb{R}^{4}$, such that Higgs field and potential
can be generated by the exotic background, the resulting YM theory
should become a kind of effective APEGT theory. In this way the agreement
of YM theory without Higgs, with the effects of the exotic background
where Higgs field comes from, can be achieved. That is why we claim
that pure $SU(2)$ YM on $e$ agrees with APEGT on $\mathbb{R}^{4}$
in the limit where Higgs field and magnetic monopoles decouple from
exotic 4-geometry. One way to see this result is to show that the
effective Higgs field (the scalar St\"uckelberg field in APEGT) and
magnetic monopole current, are generated by the exotic geometry of
$e$. Next, given the bare possibility that YM theory can be formulated
on a smooth $e$ we arrive at the conclusion that YM on $e$ should
be twisted like APEGT. The effects of exotic geometry of $e$ are
represented by the appearance of the Higgs field and the magnetic
current. On the other hand, APEGT directly predicts confinement, so
we infer that the geometry of $e$ causes the confinement in YM on
$e$.

APEGT is a version of an effective theory derived from maximal abelian
projection or gauge in YM theory. The MAG in case of $SU(2)$ YM theory
is performed via the projection onto the maximal weight $\phi_{0}^{3}=\frac{\sigma_{3}}{2}$
and in the representation where it is diagonal. The generators of
$SU(2)$ are as usual $T^{a}=\frac{1}{2}\sigma^{a},a=1,2,3$ and $\sigma^{a}$
are Pauli matrices. In a general case, an abelian projection on a
field $\phi=\overset{\to}{\phi}\cdot\overset{\to}{\sigma}=\phi^{a}\sigma^{a}$
in the adjoint of $SU(2)$, is defined via the following procedure.
Gauge invariant projected field strength (the non-abelian 't Hooft
tensor) is given by:
\[
F_{\mu\nu}^{(\phi)}=\hat{\phi}\cdot F_{\mu\nu}-\frac{1}{g}\hat{\phi}\cdot(D_{\mu}\hat{\phi}\wedge D_{\nu}\hat{\phi})=\hat{\phi}^{a}F_{\mu\nu}^{a}-\frac{1}{g}\hat{\phi}^{a}(D_{\mu}\hat{\phi}^{a}\wedge D_{\nu}\hat{\phi}^{a})
\]
where $\hat{\phi}=\frac{\overset{\to}{\phi}}{|\overset{\to}{\phi}|}$
is the vector of $\phi$ in the iso-space and $D_{\mu}\hat{\phi}=(\partial_{\mu}-gA_{\mu}\wedge)\hat{\phi}$.
The projected field strength $F_{\mu\nu}^{(\phi)}$ is thus rewritten
as: 
\[
F_{\mu\nu}^{(\phi)}=\partial_{\mu}(\hat{\phi}^{a}A_{\nu}^{a})-\partial_{\nu}(\hat{\phi}^{a}A_{\mu}^{a})-\frac{1}{g}\hat{\phi}^{a}(\partial_{\mu}\hat{\phi}^{a}\wedge\partial_{\nu}\hat{\phi}^{a})\,.
\]
Fixing $(\hat{\phi}^{a})=(0,0,1)$ the result is 
\[
F_{\mu\nu}^{(3)}=\partial_{\mu}A_{\nu}^{3}-\partial_{\nu}A_{\mu}^{3}
\]
which is the abelian projection on $\hat{\phi}$ and possesses the
residual $U(1)$ symmetry given by the rotation around $\hat{\phi}$.
Magnetic monopoles can exist whenever there is a non-zero dual current
$j_{\nu}=\partial^{\mu}{^{\star}F}_{\mu\nu}$.

In the $SU(2)$ YM theory the non-zero current $J_{\mu}=D_{\nu}{^{\star}F_{\mu\nu}}$
where $^{\star}F_{\mu\nu}=\frac{1}{2}\epsilon_{\mu\nu\rho\sigma}F_{\rho\sigma}$
is the ordinary dual field strength of the gauge potential $A$, directly
violates the non-abelian Bianchi identity, i.e. $D_{\nu}{^{\star}F_{\mu\nu}}\neq0$.
As shown in Ref. \cite{DiGiacomo-2010} the projected abelian magnetic
current $j_{\mu}$ is proportional to $J_{\mu}$, i.e. 
\[
\partial_{\mu}{^{\star}F_{\mu\nu}^{(\phi)a}}={\rm Tr}(\phi^{a}J_{\nu})
\]
where now ${^{\star}F_{\mu\nu}^{(\phi)a}}$ is the dual 't Hooft tensor.
As a result, the violation of the Bianchi identity is the sufficient
and necessary condition for the appearance of the magnetic current
in solutions of YMH theory. The abelian 't Hooft tensor $f_{\mu\nu}$
is assigned to the abelian projection $\phi^{3}$ as the abelian field
strength of the residual $U(1)$ gauge group.

To understand more clearly the reason behind the confinement in $SU(2)$
YM theory on exotic $\mathbb{R}^{4}$ one looks at the singularity
of $A_{\mu}^{3}(x)$ which is the source for magnetic currents after
the abelian projection as above. Then, one can make use of the (double-charged
abelian) Wilson loop and it is shown that the usual area law holds
for them, i.e. $W_{2}(C)=\left<Pe^{2i\oint_{C}dx^{\mu}A_{\mu}^{3}\frac{\sigma_{3}}{2}}\right>$
and $W_{2}(C)\sim e^{-{\rm Area}(C)}$.

The abelian-projected effective gauge theory of Kondo is formed when
in pure, say $SU(2)$ YM theory on $\mathbb{R}^{4}$, the maximal
abelian projection is performed and one integrates out the off-diagonal
fields. Thus, one chooses the maximal abelian gauge leading to the
confinement in this effective theory. As we saw MAG relies on the
choice of the gauge-dependent quantity $Z(x)=Z^{a}(x)T^{a}$ which
transforms in the adjoint representation of $SU(2)$, i.e. $Z'(x)=U(x)Z(x)U^{\dag}(x)$.
Next, one diagonalizes $Z(x)$ by the gauge rotation, so that the
eigenvalues appear as $\lambda_{i}(x),i=1,2$ on the diagonal of the
transformed $Z(x)={\rm diag}(\lambda_{1}(x),\lambda_{2}(x))$.

Whenever $\lambda_{i}(x)=\lambda_{j}(x),i\neq j$ for $x\in\mathbb{R}^{4}$
the singularity appears which is the 't Hooft magnetic monopole solution.
In fact, the gauges leaving the procedure, i.e. $Z(x)$, invariant
reduces now to the $U(x)={\rm diag}(e^{i\theta_{1}},e^{i\theta_{2}})$
such that $\theta_{1}+\theta_{2}=0$. But these gauges reduce $SU(2)$
YM theory to the $U(1)$ abelian theory. More explicitly, according
to the above prescription, the gauge vector potential is decomposed
into the diagonal $U(1)$ residual part (the maximal torus) and the
off-diagonal $SU(2)/U(1)$: 
\[
A_{\mu}(x)=\sum_{a=1}^{3}A_{\mu}^{a}(x)T^{a}=a_{\mu}(x)T^{3}+\sum_{a=1}^{2}A_{\mu}^{a}(x)T^{a}\,.
\]
One decomposes the usual field strength $F_{\mu\nu}^{a}T^{a}=(\partial_{\mu}A_{\nu}^{a}(x)-\partial_{\nu}A_{\mu}^{a}(x)-i[A_{\mu}^{a},A_{\nu}^{a}])T^{a}$
as: 
\[
F_{\mu\nu}^{a}T^{a}=[f_{\mu\nu}(x)+C_{\mu\nu}(x)]T^{3}+S_{\mu\nu}^{2}T^{2}+S_{\mu\nu}^{1}T^{1}
\]
where again the summation on $a=1,2,3$ is assumed. Here \cite{Kondo-1997}:
\[
\begin{array}{c}
f_{\mu\nu}(x)=\partial_{\mu}a_{\nu}(x)-\partial_{\nu}a_{\mu}(x)\,,\\[5pt]
S_{\mu\nu}^{a}(x)=\partial_{\mu}\delta^{ab}A_{\nu}^{b}-\epsilon^{ab3}a_{\mu}A_{\nu}^{b}-[\partial_{\nu}\delta^{ab}A_{\mu}^{b}-\epsilon^{ab3}a_{\nu}A_{\mu}^{b}]\,,a,b=1,2\\[5pt]
C_{\mu\nu}(x)T^{3}=-i[A_{\mu}(x),A_{\nu}]\,,
\end{array}
\]
so that the diagonal part reads: 
\[
F_{\mu\nu}^{3}=f_{\mu\nu}+A_{\mu}^{1}A_{\nu}^{2}-A_{\nu}^{2}A_{\mu}^{1}\,.
\]
The pure YM action (\ref{YM-4}) can be rewritten in terms of the
diagonal field, as: 
\[
S_{YM}=-\frac{1}{4g^{2}}\int d^{4}x\,[(f_{\mu\nu}+C_{\mu\nu})^{2}+(S_{\mu\nu}^{a})^{2}]\ \,.
\]
One introduces the dual tensor field to the diagonal $F_{\mu\nu}^{3}$
as: 
\[
B_{\mu\nu}=\frac{1}{2}\epsilon^{\mu\nu\rho\sigma}F_{\rho\sigma}^{3}=\frac{1}{2}\epsilon^{\mu\nu\rho\sigma}(f_{\rho\sigma}+C_{\rho\sigma})\,
\]
and its Hodge decomposition gives: 
\begin{equation}
B_{\mu\nu}=b_{\mu\nu}+\tilde{\chi}_{\mu\nu},\; b_{\mu\nu}=\partial_{\mu}b_{\nu}-\partial_{\nu}b_{\mu},\;\tilde{\chi}_{\mu\nu}=\frac{1}{2}\epsilon_{\mu\nu\rho\sigma}(\partial^{\rho}\chi^{\sigma}-\partial^{\sigma}\chi^{\rho})\;.\label{Hodge}
\end{equation}
Introducing the dual $\tilde{f}_{\mu\nu}=\frac{1}{2}\epsilon_{\mu\nu\rho\sigma}f^{\rho\sigma}$
the (gauge fixed) YM action reads: 
\[
\overline{S}_{YM}=\int d^{4}x\left[-z_{a}\frac{1}{g^{2}}f_{\mu\nu}f^{\mu\nu}-z_{b}\cdot g^{2}(b_{\mu\nu}b^{\mu\nu}+\tilde{\chi}_{\mu\nu}\tilde{\chi}^{\mu\nu})+\frac{1}{2}z_{c}b_{\mu\nu}\tilde{f}_{\mu\nu}+\frac{1}{2}z_{c}\chi_{\mu\nu}f_{\mu\nu}+...\right]
\]
which, after integrating out $\chi$ and neglecting the ghost self-interaction
and some higher derivatives, becomes the effective theory written
in terms of the abelian $a_{\mu}$ and the dual $b_{\mu}$: 
\[
S=\int d^{4}x\left[-z_{a}\frac{1}{g^{2}}f_{\mu\nu}f^{\mu\nu}+i\overline{c}^{a}D_{\mu}^{ac}[a]D_{\mu}^{cb}[a]c^{b}-z_{b}\cdot g^{2}b_{\mu\nu}b^{\mu\nu}-z_{c}b_{\mu}k^{\mu}\right]\,.
\]
Here $D_{\mu}[a]^{ab}=\partial\delta^{ab}-\epsilon^{ab3}a_{\mu}$
as before and $c^{a}$, $\overline{c}^{a}$ are ghost, anti-ghost
fields due to the gauge fixing functional as in \cite{Kondo-1997}.
We do not make use of the explicit form of the expressions $z_{a},z_{b},z_{c}$
appearing here and the interested reader can find them in \cite{Kondo-1997}.
The crucial point is the presence of the magnetic current $k_{\mu}=\partial^{\nu}\tilde{f}_{\mu\nu}$
of the dual field strength $\tilde{f}_{\mu\nu}$. The confinement
via the dual Meissner effect appears whenever the diagonal $a_{\mu}$
is singular - in this case the magnetic current $k_{\mu}$ is non-zero
and it couples to the dual $b_{\mu}$. We propose here the purely
geometric way of generating the singularity of the diagonal potential
$a_{\mu}$. Let us suppose that the YM theory is formulated on an
\emph{exotic} $\mathbb{R}^{4}$ in a specific way. Namely, the diagonal
components of the dual vector potential $A_{\mu}$ propagates on exotic
geometry while the off-diagonal ones are not sensitive to the exoticness,
hence propagate on the standard $\mathbb{R}^{4}$. Thus $a_{\mu}$
propagates on exotic $\mathbb{R}^{4}$ either. Next, let us make use
of the fundamental difference between standard and exotic smoothness
on $\mathbb{R}^{4}$: (i) \emph{Smooth functions in the standard and
exotic structures are not the same, i.e. there should exist continuous
non-differentiable functions in the standard smooth structure which
are smooth in an exotic structure on $\mathbb{R}^{4}$.} Let $A_{\mu}(x)$
be such a function. Additionally we should argue that singularities
in $A_{\mu}$ leading to the confinement are of the kind generated
by the change of the smoothness as above. To this end let us consider
the abelian projected field strength $f_{\mu\nu}$ as given directly
in terms of the singular gauge potential $A_{\mu}$: 
\begin{equation}
f_{\mu\nu}={\rm Tr}(ig[A_{\mu},A_{\nu}]T^{3})+{\rm Tr}(\frac{i}{g}U(x)[\partial_{\mu},\partial_{\nu}]U^{\dag}(x)T^{3})\label{YM-6}
\end{equation}
where $U(x)$ is a gauge transformation. The magnetic current of the
magnetic monopoles derived from $f_{\mu\nu}$ reads: $k_{\mu}=\frac{1}{2}\epsilon_{pq\mu s}\partial^{s}f_{pq}$
which is equivalent to the contribution from the first term on the
RHS of (\ref{YM-6}), i.e. $K_{\mu}=\frac{1}{2}\epsilon_{pq\mu s}\partial^{s}(g\epsilon^{ab3}A_{p}^{a}A_{q}^{b})$.
The singularity of $A$ can be developed by the abelian gauge-fixing
conditions put on the transformed gauge potential $A'_{\mu}(x)=U(x)A_{\mu}(x)U^{\dag}(x)+\frac{i}{g}U(x)\partial_{\mu}U^{\dag}(x)$.
The abelian gauge field $a_{\mu}={\rm Tr}(T^{3}A_{\mu})$ corresponding
to $A$ develops a singularity under the gauge transformation $U(x)$,
too. The corresponding transformation for the field strength reads:
\[
F'_{\mu\nu}(x)=\partial_{\mu}A'_{\nu}(x)-\partial_{\nu}A'_{\mu}(x)-ig[A'_{\mu}(x),A'_{\nu}(x)]
\]
such that the corresponding abelian field strength is: 
\[
f_{\mu\nu}=\partial_{\mu}a'_{\nu}-\partial_{\nu}a'_{\mu}={\rm Tr}[T^{3}(U(x)F_{\mu\nu}U^{\dag}(x)+ig[A'_{\mu},A'_{\nu}])]\,.
\]
For the magnetic current $k_{\mu}=\frac{1}{2}\epsilon_{\mu\nu pr}\partial^{\nu}f^{pr}$,
only when $U(x)$ is singular the term $\frac{i}{g}U\partial_{\mu}U^{\dag}$
in the expression for $A'_{\mu}(x)$, gives the non-vanishing contribution
to $k_{\mu}$. As shown in Ref. \cite{Kondo-1997} the contributions
to $k_{\mu}(x)$ derived from this singular $A_{\mu}(x)$ are of two
kinds: first derived from the first term in the RHS of (\ref{YM-6})
and it is a magnetic monopole sitting at $\overset{\to}{r}=0$, and
the second, corresponding to the second term in the RHS in (\ref{YM-6}),
is the 'Dirac string'. Both contributions have distributional character.
Namely, the calculation of the contribution from the magnetic monopole
part from the singular $A_{\mu}(x)$, gives the magnetic current as
\cite{Kondo-1997}: 
\[
k_{\mu}=\frac{4\pi}{g}\delta_{\mu0}\delta^{(3)}(x)\,.
\]
Precisely this kind of contributions allows for the appearance of
the effects of magnetic monopoles, hence also confinement, in YM theory.
Again, the confinement can be detected via Wilson loops observables
and it is shown that the usual area law holds true also in this case.

One could wonder whether such singularities (contributions) might
appear as the result of the change of smoothness on the background
$\mathbb{R}^{4}$. From general considerations on distribution theory,
one infers that every tame distribution is derivable from some continuous
functions by taking their distributional derivatives and combinations
thereof. Given a tempered distribution $h$ one can always find a
continuous slowly increasing function $H$, such that $h=D^{|\alpha|}H$
for some multiindex $|\alpha|$. Similar results hold true also for
more general classes of distributions. So, starting with continuous
functions and performing distributional derivatives gives rise to
general distributions.

Turning to exotic smooth structures on $\mathbb{R}^{4}$ which would
be different non-diffeomorphic ones, merely continuous functions in
one structure which are smoothly differentiable in the other, have
to exist. Otherwise, the structures would be diffeomorphic. The results
of Ref. \cite{Krol:04b} show that a 'general singularity' emerges
as some distribution in the process of changing the smoothness on
$\mathbb{R}^{4}$ to the standard one. One finds a continuous function which
is smoothly differentiable as exotic smooth function, and by 
the distributional differentiation, after (possibly)
infinite many differentiations, the suitable singularity emerges. Moreover, there are infinite (continuum)
many different small smooth $\mathbb{R}^{4}$s in the fixed radial
family, hence from the general point of view there is a good chance
to model the required distributional singularity by some exotic structure
from the family.

However, there are also direct reasons explaining why the non-zero
magnetic current appears in YM theory on exotic $\mathbb{R}^{4}$
after the change of smoothness into the standard one. Let us consider
$a_{\mu}(x)$ and $A_{\mu}(x)$ as propagating on some exotic $\mathbb{R}^{4}$
from the fixed radial family. As we know the characteristic feature
of all such small $\mathbb{R}^{4}$ is that they determine codimension-one
foliations of $S^{3}$ with non-zero GV invariant related to the radius
of the family. We make use of this non-zero GV invariant directly:
\[
{\rm GV}\neq0\;{\rm means\; that}\;0\neq[d\theta\wedge\theta]\in H^{3}(S^{3},\mathbb{R})
\]
where $\theta$ is the one form derived from the integrability condition
of the 1-form $\omega$ for any codimension-one foliation, i.e. $d\omega\wedge\omega=0$
and $d\theta=-\theta\wedge\omega$. The differential forms here can
be considered as defined on the separating 3-manifold $Y_{n}$ or
on 3-sphere $S^{3}$ from which $Y_{n}$ is obtained via the surgery
alone some knot \cite{AsselmeyerKrol2009}. Such realization of the
non-zero GV from exotic 4-space is a global effect derived from an
exotic $\mathbb{R}^{4}$. Extending 1-forms $\theta$ over the 4-region
still has to give nontrivial GV number when restricting the form on
$Y_{n}$ and integrating over it.

This observation suggests that the restricted abelian vector potential
$a_{\mu}^{3}(x)_{|Y_{n}}$ and after the projection on the standard
$\mathbb{R}^{4}$ one can take a certain 1-form proportional to $\theta$.
However, $Y_{n}$ is embedded in the exotic $\mathbb{R}^{4}$ and
\[
0\neq{\rm GV}_{Y_{n}}\in H^{3}(Y_{n},\mathbb{R}),\;{\rm so\; that}\; d\theta\sim f_{\mu\nu\,|Y_{n}}\;{\rm and}\;0\neq\epsilon_{\nu pr}\partial^{\nu}f_{|Y_{n}}^{pr}\in H^{3}(Y_{n},\mathbb{R})\,.
\]
As the result 
\[
k_{\mu|Y_{n}}=\frac{1}{2}\epsilon_{\mu\nu pr}\partial^{\nu}f_{|Y_{n}}^{pr}\neq0\,.
\]
Moreover, $S^{3}$ is embedded in the standard $\mathbb{R}^{4}$ and
one can always choose the codimension-one foliation of $S^{3}$ such
that $0\neq{\rm GV}_{S^{3}}={\rm GV}_{Y_{n}}$ and work with $H^{3}(S^{3},\mathbb{R})$
in the standard $\mathbb{R}^{4}$ \cite{AsselmeyerKrol2009}. That
is why 
\[
{\rm GV}_{S^{3}}[S^{3}]\sim\frac{1}{2}\int_{S^{3}}dv\,\epsilon_{\mu\nu pr}\partial^{\nu}f_{|S^{3}}^{pr}\neq0\,.
\]
The extension of the nontrivial class (non-exact 3-form) in $H^{3}(S^{3},\mathbb{R})$
to a 4-dimensional region gives rise to the singular expressions for
the magnetic current and vector potential. On the other hand, a given
standard $\mathbb{R}^{4}$ and a vector potential $a_{\mu}(x)$ propagating
on it, one has ${\rm GV}=0$ and $k_{\mu}=0$, i.e. no nontrivial
magnetic current exists for these global non-singular vector bosons.

Consider in the following that the singularity in $a_{\mu}$ or $A_{\mu}$
is the result of changing the smoothness from exotic to the standard
$\mathbb{R}^{4}$. The diagonal vector fields propagate on exotic
smooth geometry on $\mathbb{R}^{4}$ and they are exotic smooth and become
singular when the smoothness is changed into standard one. As the
result, the non-trivial magnetic current emerges. The theory shows
confinement when additionally the effects are required to be compatible
with the non-trivial effective Higgs potential, hence $v\neq0$, derived
from the exotic handle-body. To explain it qualitatively let us note
that in the APEGT the effective abelian magnetic monopole current,
reads:
\[
K^{\mu}=\frac{1}{2}\epsilon^{\mu\nu\rho\sigma}\partial_{\nu}(\epsilon^{ab3}A_{\rho}^{a}A_{\sigma}^{b})\,,
\]
as we observed already and which is the contribution of the first
term in the RHS of (\ref{YM-6}). The dual magnetic field $b_{\mu}$
stays massless whenever the magnetic $U(1)$ symmetry is not broken.
This result follows from extracting the $b_{\mu}$-dependent parts
of the effective action of APEGT \cite{Kondo-1997}. The non-zero
mass for $b_{\mu}$ breaks the $U(1)$ symmetry so that the compact
correlation function for the magnetic current becomes 
\begin{equation}
\left<K_{\mu}K_{\nu}\right>=g^{2}\delta_{\mu\nu}\delta^{(4)}(x-y)f(x)+...\label{cond}
\end{equation}
and the mass term for the dual field $b_{\mu}$ appears in the action
for this field, as
\begin{equation}
S[b]=\int d^{4}x[\frac{-1}{4}b_{\mu\nu}b^{\mu\nu}+\frac{1}{2}g^{2}f(x)b_{\mu}(x)b_{\mu}(x)+....]\label{S}
\end{equation}
where $f(x)=m_{b}^{2}$ is the square of the mass, and $b_{\mu\nu}=\partial_{\mu}b_{\nu}-\partial_{\nu}b_{\mu}$
as in (\ref{Hodge}). This is the dual Meissner effect for the APEGT
without the Higgs field where the dual gauge field acquires the non-zero
mass due to the condensation of magnetic monopoles as in (\ref{cond}).

However, the 'Higgs' scalar $\phi(x)$ can be reintroduced into the
theory as the so called St\"uckelberg field. Again following \cite{Kondo-1997},
taking $\phi(x)=\frac{m_{b}}{\sqrt{2}}e^{i\theta(x)}$, the mass term
$\frac{1}{2}m_{b}^{2}b_{\mu}(x)b_{\mu}(x)$ is rewritten as 
\[
\frac{1}{2}m_{b}^{2}(b_{\mu}(x)-\partial_{\mu}\theta(x))^{2}=|(\partial_{\mu}-ib_{\mu}(x))\phi(x)|^{2}
\]
which is $U(1)$-invariant. It appears that our dual abelian gauge
theory given by $S[b]$ in (\ref{S}) is now equivalent to the dual
Ginzburg-Landau (GL) theory, i.e. the dual abelian Higgs theory (\ref{LG-1}):
\[
S[b]=\int d^{4}x\left[\frac{-1}{4}b_{\mu\nu}b^{\mu\nu}+|(\partial_{\mu}-ig^{-1}b_{\mu}(x))\phi(x)|^{2}+\lambda(|\phi(x)|^{2}-\frac{m_{b}^{2}}{2})^{2}+...\right]\,.
\]
This is best seen by taking the so called London limit in this GL
theory where also the coupling constant is reversed, i.e. $\lim_{\lambda\to\infty}\lambda(|\phi|^{2}-\frac{m_{b}^{2}}{2})^{2}$
and $g\to\frac{1}{g}$. Now, in analogy with the abelian YMH theory,
the deconfining phase corresponds to the $m_{b}=0$ and the confinement
appears whenever the minimum of the potential $V(\phi)$ is non-zero,
hence the $m_{b}\neq0$. This last case, however, was derived without
the Higgs scalar in $SU(2)$ YM theory on an exotic $\mathbb{R}^{4}$.
Besides, it is seen that the Higgs potential generated from the exotic
handle-body appears as the dual potential in the abelian YMH theory
derived from the dual APEGT theory. This result has its reasoning
in the requirement that magnetic monopoles effects from exotic $\mathbb{R}^{4}$
agree with the Higgs potential emerging from the exotic handle-body.
Certainly, one can detect the confinement also by the suitable Wilson
line operators in APEGT \cite{Kondo-1997}.

Therefore we obtain the geometric mechanism for the confinement in
$SU(2)$ YM theory on certain small exotic $\mathbb{R}^{4}$s from
the fixed radial family. The APEGT can be considered as the effective
theory matching with the exotic 4-geometry and exhibiting the confinement
when YM theory is formulated, in the suitable way presented here,
on an exotic $\mathbb{R}^{4}$. For the smooth fields propagating
in the standard smooth structure on $\mathbb{R}^{4}$ no geometric
singularity will appear, hence no geometric magnetic monopoles for
these vector bosons exist and the theory do not show confinement.

\section{Discussion}

In this paper we show that certain differential structures on $\mathbb{R}^{4}$,
which are nonstandard smoothness structures, can push forward our
understanding of the confining phase of 4-d $SU(2)$ YM theory. As
a rule the structures are grouped in the DeMichelis-Freedman type
radial family of smooth exotic $\mathbb{R}^{4}$s. Since there is
still disagreement about the pure YM theory structure of the vacuum
and the origin of the confinement from the asymptotically free infrared
part of the theory, our proposal gives some new geometric perspective
on these problems. The scenario advocated here can be described as
follows. In the infrared limit all gluons propagate on the standard
4-geometry, while in sufficiently low energies some (diagonal, dual)
modes propagate on an exotic $\mathbb{R}^{4}$. This is the geometric
reason for breaking the non-abelian gauge symmetry and performing
the abelian projection. Thus, we find the reason for the abelian projection rather
in the outside geometry of the background, than in
the YM theory itself. 

Further breaking of the remnant dual $U(1)$
symmetry is also explained via fake smoothness of $\mathbb{R}^{4}$.
The exotic geometry decouples from the theory in sufficiently high
energies and the deconfining phase is reached. This specific relation
(interaction) of geometry and YM theory is possible only in dimension
four.

Magnetic monopoles appear in the theory when fake 4-geometry is broken
to the standard one. Equivalently, exotic 4-geometry gives rise to
the magnetic current and Higgs potential in YM theory on $\mathbb{R}^{4}$.
Fixing the radial family of small exotic $\mathbb{R}^{4}$s gives
the codimension-1 foliations of some compact 3-manifolds: $Y_{n}$
embedded in the exotic $\mathbb{R}^{4}$, and $S^{3}$ in the standard
$\mathbb{R}^{4}$. Both foliations have the same, non-zero value of
the Godbillon-Vey (GV) invariant \cite{Asselm-Krol-2012g}. We have
shown in a direct way that the confinement in YM theories is caused
by the exoticness of open background 4-manifolds where the GV class
breaks the dual magnetic abelian symmetry. This idea is completely
new approach to the confinement in YM theories. The theory, which
effectively encompasses the effects from exotic geometry and those
of deformed YM theory, is chosen as APEGT of Kondo. Some other choices
are still possible. Recent proposal for the analytical generation
of confinement, is to reformulate YM theory in terms of new variables
and then to apply the dual Meissner effect \cite{Kondo-2011a,Kondo-2010b,DiGiacomo-2010}.

Several points require further comments. In the paper we considered
the monopoles condensation model for confinement in the $SU(2)$ YM
theory on exotic $\mathbb{R}^{4}$ where monopoles are generated by
this fake geometry. Other mechanisms for confinement, like Gribov-Zwanziger
and Dyson-Schwinger propagators, vacuum wave-functions or gluon chains
\cite{Green-2011}, are also likely to contain elements derivable
from exotic 4-geometry. Even though, they are not directly accessible
due to the entirely unknown analytical shape of the global exotic
geometry on $\mathbb{R}^{4}$, their conjectural agreement with the
proposal, would serve as a test for it. In our case, magnetic monopoles,
Higgs potential, breaking of the dual magnetic $U(1)$ symmetry and
abelian projection from non-abelian gauge theory, all these effects
are derivable from general properties of the radial family and handle-body
smooth topology of the exotic $\mathbb{R}^{4}$.

The extension of the $SU(2)$ YM theory into the more physical $SU(3)$
case and including the matter (quarks) is possible. This part of the
theory will be performed separately, however, the important ingredient
would presumably be the topology of some $S^{2}$ fibrations which
makes the $SU(3)$ case rather exceptional. Again, the propagation of
some fields (not all) in exotic geometry of the background $\mathbb{R}^{4}$
would be the reason for the appearance of the abelian projection and
confinement. Again, exotic 4-geometry of the background can break
the dual $U(1)$ magnetic symmetry. The deeper discussion of the quantum
level of YM theories on exotic $\mathbb{R}^{4}$ is also an important
challenge and can be approached from various perspectives which sheds
further light on confinement.

Wilson loops display the confinement in the YM theory. If the confinement
comes from an exotic $\mathbb{R}^{4}$, certain Wilson loops observables
could be further used for detecting the exotic smoothness on background
$\mathbb{R}^{4}$. The fundamental question which emerges here, is
whether one can propose specific observables in 'real' YM theory and
specific methods to measure them which would indicate, also experimentally,
the presence of an exotic background for $SU(3)$ YM theory with matter.
This point becomes more important because QCD non-perturbative calculations
show severe limitations from the fundamental point of view. In this
case, the connection of exotic 4-geometry with some states of 'effective'
matter, claimed some time ago, would not be only a bold formal guess
\cite{AsselmKrol11,AsselmKrol2011f}. After all, breaking the non-abelian
gauge, and the remnant magnetic, symmetries due to exotic smooth structures,
could leave specific experimental trace. To identify it properly requires,
however, a careful analysis and further insights.

\end{document}